\documentclass{emulateapj}
\usepackage{apjfonts}
\usepackage{mathptmx}
\usepackage{epsf}

\newcommand{\hdf}{HDF--N}
\newcommand{\hdfs}{HDF--S}
\newcommand{\cdfs}{CDF--S}
\newcommand{\hst}{\textit{HST}}

\newcommand{\wfv}{\hbox{$V_{606}$}}
\newcommand{\wfi}{\hbox{$I_{814}$}}
\newcommand{\acsb}{\hbox{$B_{435}$}}
\newcommand{\acsv}{\hbox{$V_{606}$}}
\newcommand{\acsi}{\hbox{$i_{775}$}}
\newcommand{\acsz}{\hbox{$z_{850}$}}

\newcommand{\ks}{\hbox{$K_s$}}
\newcommand{\AAA}{\hbox{\AA}}
\newcommand{\lya}{Lyman~$\alpha$}

\newcommand{\lsim}{\lesssim}
\newcommand{\gsim}{\gtrsim}

\newcommand{\etal}{et al.}
\newcommand{\eg}{e.g.}

\newcommand{\kmsmpc}{\hbox{km~s$^{-1}$~Mpc$^{-1}$}}

\slugcomment{Accepted for Publication in the Astrophysical Journal Letters}
\submitted{Accepted for Publication in the Astrophysical Journal Letters}

\shorttitle{EVOLUTION IN THE COLORS OF LYMAN--BREAK GALAXIES}
\shortauthors{PAPOVICH ET AL.}

\begin{document}

\title{EVOLUTION IN THE COLORS OF LYMAN--BREAK GALAXIES FROM $Z\sim 4$
  TO $Z\sim 3$\altaffilmark{1}}	

\author{\sc Casey Papovich\altaffilmark{2}, 
        Mark Dickinson\altaffilmark{3,4},
        Henry C. Ferguson\altaffilmark{3,4},
        Mauro Giavalisco\altaffilmark{3},
	Jennifer Lotz\altaffilmark{5}, \\
	Piero Madau\altaffilmark{6},
	Rafal Idzi\altaffilmark{4},
	Claudia Kretchmer\altaffilmark{4},
	Leonidas A. Moustakas\altaffilmark{3},
	Duilia F. de~Mello\altaffilmark{7}, \\
	Jonathan P. Gardner\altaffilmark{7},
	Marcia J. Rieke\altaffilmark{2},
	Rachel S. Somerville\altaffilmark{3},
	and Daniel Stern\altaffilmark{8}
}

\altaffiltext{1}{Based on observations taken with the NASA/ESA Hubble
Space Telescope, which is operated by the Association of Universities
for Research in Astronomy, Inc.\ (AURA) under NASA contract
NAS5--26555, and based on observations collected at the European
Southern  Observatory, Chile (ESO Programmes 168.A-0485,  64.O-0643,
66.A-0572, 68.A-0544).}
\altaffiltext{2}{Steward Observatory, The University
of Arizona,  933 North Cherry Avenue, Tucson, AZ 85721;
\texttt{papovich, mrieke@as.arizona.edu}} 
\altaffiltext{3}{Space Telescope
Science Institute, 3700 San Martin Drive, Baltimore, MD 21218;
\texttt{med, ferguson, mauro, leonidas, somerville@stsci.edu}}
\altaffiltext{4}{Department of Physics and Astronomy, The Johns
Hopkins University, Baltimore, MD 21218;
\texttt{idzi, claudiak@stsci.edu}}
\altaffiltext{5}{Department of Physics, University of California at
Santa Cruz, CA 95064; \texttt{jlotz@scipp.ucsc.edu}}
\altaffiltext{6}{Department of Astronomy and Astrophysics, University
of California at Santa Cruz, Santa Cruz, CA 95064;
\texttt{pmadau@ucolick.org}}
\altaffiltext{7}{Laboratory for Astronomy and Solar Physics, Code 681,
NASA GSFC, Greenbelt, MD 20771;
\texttt{dui\-lia@ipa\-ne\-ma.gsfc.nasa.gov}, 
\texttt{gardner@harmony.gsfc.nasa.gov}}
\altaffiltext{8}{Jet Propulsion Laboratory, California Institute of
Technology, Mail Stop 169-506, Pasadena, CA 91109;
\texttt{stern@zwolfkinder.jpl.nasa.gov}}


\begin{abstract}

The integrated colors of distant galaxies provide a means for
interpreting the properties of their stellar content.  Here, we use
rest--frame UV--to--optical colors to constrain the spectral--energy
distributions and stellar populations of color--selected, $B$--dropout
galaxies at $z\sim 4$ in the \textit{Great Observatories Origins Deep
Survey}.  We combine the ACS data with ground--based near--infrared
images, which extend the coverage of galaxies at $z\sim 4$ to the
rest--frame $B$--band.  We observe a color--magnitude trend in the
rest--frame $m(\mathrm{UV}) - B$ versus $B$ diagram for the $z\sim 4$
galaxies that has a fairly well--defined ``blue--envelope'', and is
strikingly similar to that of color--selected, $U$--dropout galaxies
at $z\sim 3$.  We also find that although the co-moving luminosity
density at rest--frame UV wavelengths (1600\AA) is roughly comparable
at $z\sim 3$ and $z\sim 4$, the luminosity density at rest--frame
optical wavelengths increases by about one--third from $z\sim 4$ to
$z\sim 3$.  Although the star--formation histories of individual
galaxies may involve complex and stochastic events, the evolution in
the global luminosity density of the UV--bright galaxy population
corresponds to an average star--formation history with a
star--formation rate that is constant or increasing over these
redshifts. This suggests that the evolution in the luminosity density
corresponds to an increase in the stellar--mass density of $\gsim
33$\%.

\end{abstract}
 
\keywords{
early universe --- 
cosmology: observations --- 
galaxies: evolution --- 
galaxies: formation --- 
galaxies: high--redshift ---
galaxies: photometry
}


\section{Introduction}

Current investigations of high--redshift ($z \gsim 2$) galaxies have
been focusing on the properties of these objects as a global
population.  Many surveys identify these galaxies by their strong
emission at rest--frame UV wavelengths (observed--frame optical) and
spectral breaks at \lya\ and the Lyman limit (so--called Lyman--break
galaxies [LBGs]; \eg, Giavalisco 2002). These
galaxies are generally dominated by the light from OB stars, and have
properties that are similar to local starburst galaxies
\citep[\eg,][]{sha03}.  Near--infrared (NIR) photometry of galaxies at
$z \gsim 2$ extends the observations to rest--frame optical
wavelengths, probing the light from A-- and later--type stars. Several
studies have used NIR observations to constrain the properties of the
stellar populations of $z\sim 2-3$ galaxies (\eg, Sawicki \& Yee 1998;
Papovich, Dickinson, \& Ferguson 2001; Shapley \etal\ 2001; Labb\'e
\etal\ 2003; Franx \etal\ 2003), and to estimate the evolution of the
global stellar--mass density for $0 < z \lsim 3$ \citep[\eg,][]{dic03}.

At present, some of the constraints on the parameters of the galaxies'
stellar--population models are uncertain by more than an order of
magnitude (Papovich \etal\ 2001; Shapley et al.\ 2001).  Even so, the
stellar--population ages and star--formation histories of the models
have broad implications for galaxy evolution at higher redshifts (see
Ferguson, Dickinson, \& Papovich 2002).  The galaxies'
spectral--energy distributions (SEDs) contain the
integrated record of their past and current star formation.  Thus,
comparing galaxy SEDs at different redshifts allows us to improve the
constraints on the star--formation histories of these galaxies.

In this \textit{Letter}, we study galaxies at
$z\sim 4$ selected from deep imaging with the Advanced Camera for
Surveys (ACS) onboard the \textit{Hubble Space Telescope} (\hst) and
augmented with NIR observations from the ground (\S~2), and we compare
these to similar rest--frame colors for color--selected galaxies at
$z\sim 3$ from the Hubble Deep Fields, North and South (\hdf\ and --S;
\S~3).  We then discuss the SEDs of the luminosity density generated
by these galaxies, and we consider the implications on the galaxies'
star--formation histories (\S~4).  Throughout this \textit{Letter}, we
use a flat cosmology with $\Omega_m=0.3$, $\Omega_\Lambda = 0.7$, a
Hubble constant of 70~\kmsmpc, and we use AB magnitudes,
$m_\mathrm{AB} = -48.6 -
2.5\log(f_\nu/\mathrm{1\;erg\;s^{-1}\;cm^{-2}\;Hz^{-1}})$.


\section{The Observations and Galaxy Sample\label{section:data}}

\begin{figure*}
\epsscale{1.0}
\plottwo{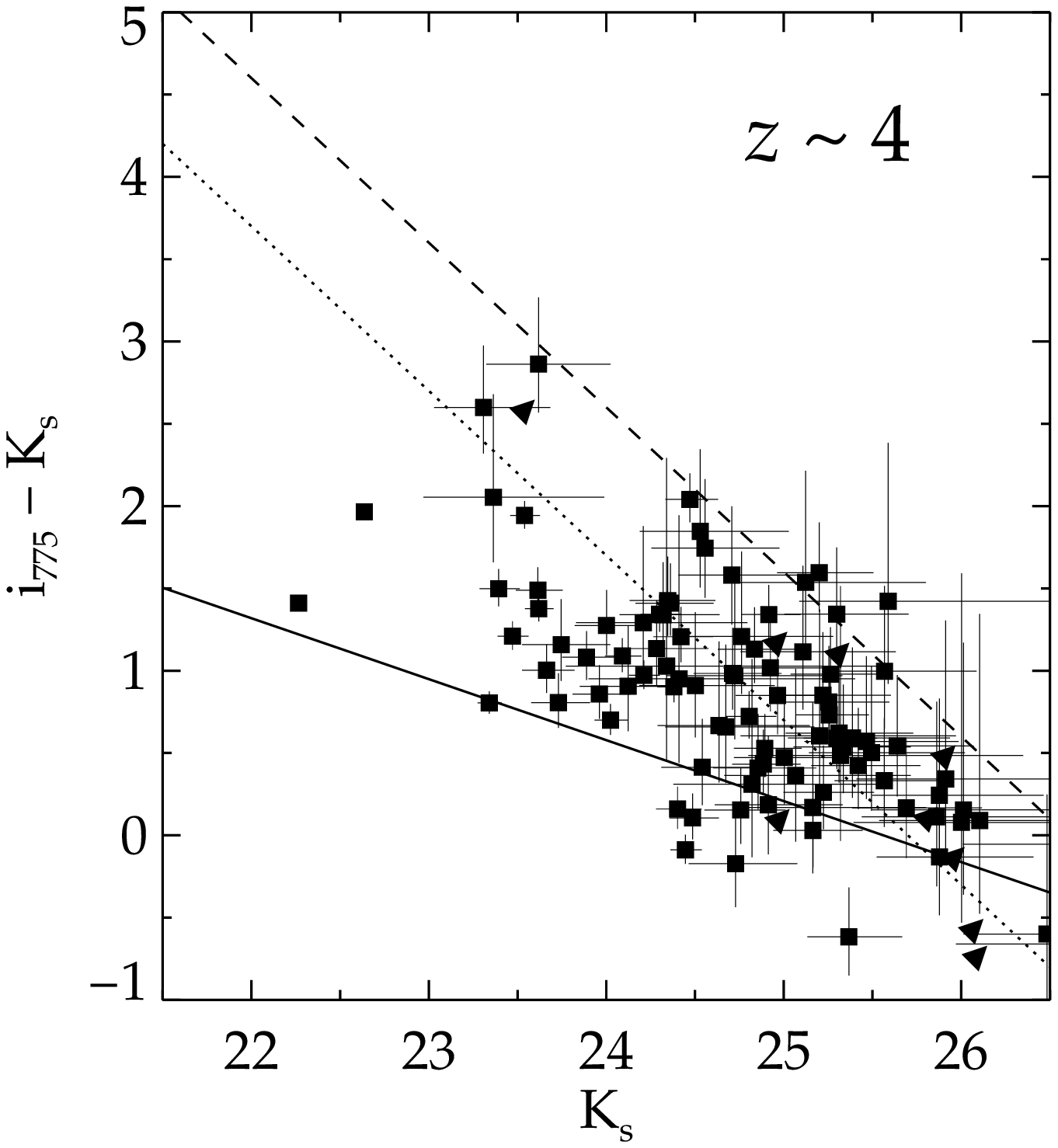}{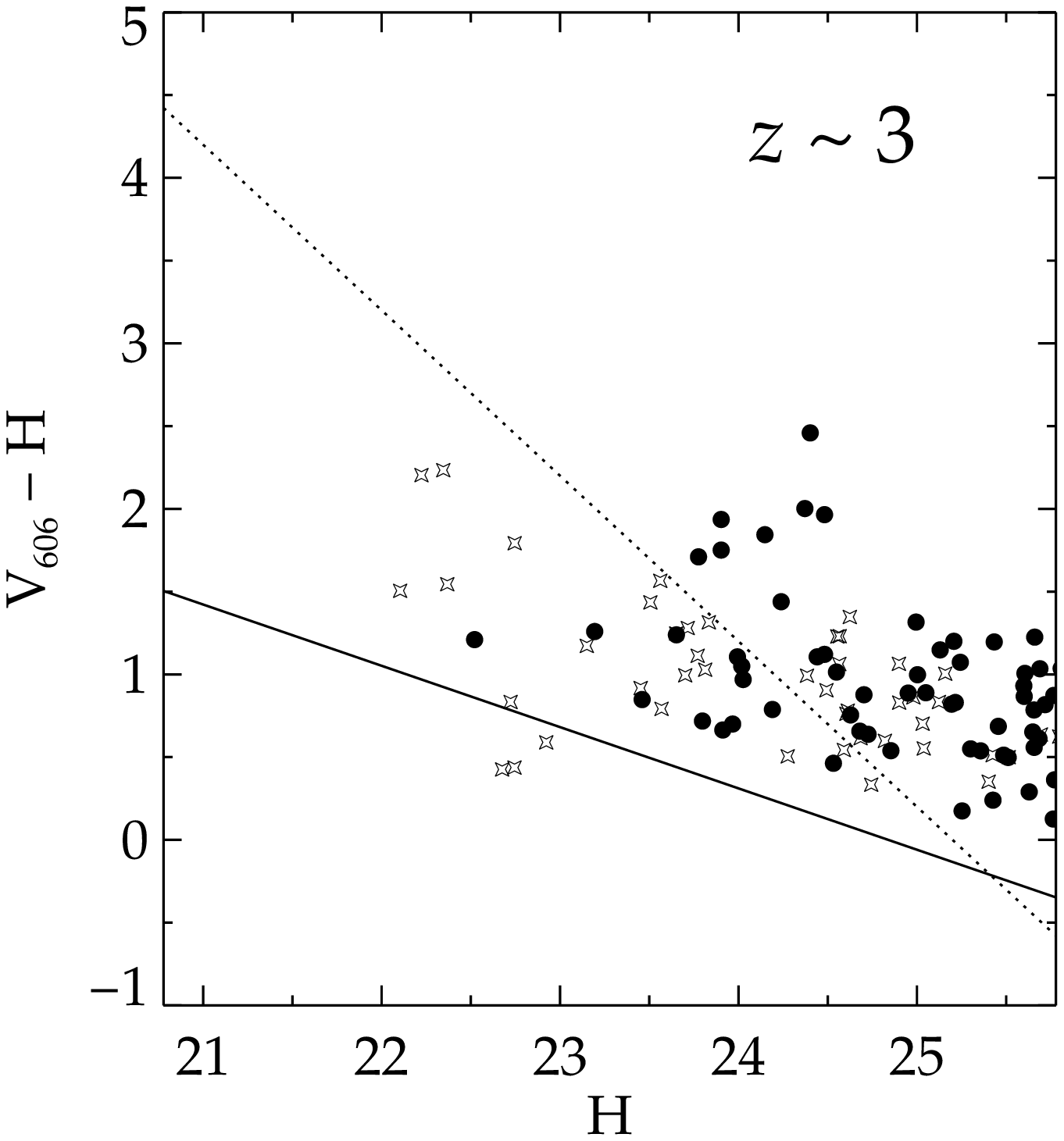}
\epsscale{1.0}
\caption{Rest--frame UV--optical color--magnitude diagrams for
  color--selected high--redshift galaxies.  \textit{Left}: Panel shows
  the $i_{775} - K_s$ versus $K_s$ color--magnitude diagram for the
  $B$--dropout galaxies selected from the GOODS CDF--S data
  (\textit{squares}; \textit{triangles} denote $1\sigma$  limits).
  These filters correspond to rest--frame $1600$-- and $4400$--\AA\
  colors.  The dashed line illustrates the approximate \acsz--band
  detection limit (for $\acsi - \acsz \approx 0$).  \textit{Right}:
  Panel shows the $V_{606} - H$ versus $H$ color--magnitude diagram
  for $U$--dropout galaxies selected from the WFPC2 data for the \hdf\
  (\textit{filled circles}) and HDF--S (\textit{open stars}).   The
  abscissa of each panel has been adjusted to show an approximately
  equal range of rest--frame $B$--band absolute magnitudes.  The
  dotted line shows the magnitude limit ($m^\ast + 1$~mag) used to
  define samples that are used to derive the luminosity
  densities in \S~3.  The solid line in each panel indicates the
  fiducial `blue--envelope'.
  \label{fig:cmd}}
\end{figure*}

The \hst\ data used in this letter stem from the first three epochs of
ACS imaging of the Chandra Deep Field South (\cdfs) as part of the
Great Observatories Origins Deep Survey (GOODS) program.  These data
provide imaging in the \acsb, \acsv, \acsi, and \acsz--bands covering
a field of view of $10\arcmin \times 16.5\arcmin$.  The dataset, its
reduction procedures, and object cataloging are described in
\citet{gia03a}.

The ACS data only probe rest--frame UV wavelengths for galaxies at
$z\gsim 1.2$.  To study the rest--frame optical light from higher
redshift galaxies we combined the ACS data with VLT/ISAAC imaging in
the $JH\ks$ bands obtained as part of GOODS \citep{gia03a}. Currently,
we use eight individual ISAAC tiles covering a total area of
50~arcmin$^2$.

We have analyzed the ISAAC $JH\ks$ images to extract optimal
photometry matched to the ACS \acsz--selected catalog.  In brief, we
use the \acsz--band data to create two--dimensional templates for each
object, convolve these to match the point--spread function (PSF) of
the ground--based data, and scale and fit the convolved
templates to the $JH\ks$--band images to extract object fluxes (see
C.\ Papovich \etal\ \textit{in preparation}; Fern\'andez-Soto,
Lanzetta, \& Yahil 1999).  The advantage of this method as applied to
faint galaxies is that it reduces concerns about PSF-- and
aperture--matching effects on the relative photometry and permits
deblending of objects partially merged by the terrestrial seeing.

We identified high--redshift galaxies whose SEDs reveal a spectral
discontinuity at the observed Lyman limit, which arises from
\ion{H}{1} absorption systems along the line of sight
\citep[\eg,][]{mad95}.  We selected ``$B$--dropout'' galaxies with,
$\acsb - \acsv \ge 1.1 + (\acsv - \acsz)$, $\acsb - \acsv \ge 1.1$,
and $\acsv - \acsz \le 1.6$ \citep[see][]{gia03b}.  Using these
colors, we are able to reject most lower--redshift interlopers while
detecting galaxies at $z\sim 4$ with bright rest--frame UV spectra
(which is indicative of ongoing star formation; Leitherer \etal\
1999).  Indeed, using the $\sim 1400$ spectroscopic redshifts
available for GOODS objects, we observe no contamination from stars or
galaxies at $z < 3$.  Our modeling indicates that these criteria
identify galaxies with an expected redshift distribution that peaks at
$\bar{z} = 3.9$ and tapers to higher and lower redshifts, with 50\%
completeness limits of $3.4 \lsim z \lsim 4.5$.


\section{UV--Optical Colors of High--Redshift Galaxies}

In the left panel of figure~\ref{fig:cmd}, we show the $\acsi - \ks$
versus $\ks$ color--magnitude diagram for the $B$--dropout galaxies.
For $z\sim 4$, these colors correspond approximately to rest--frame
$m(1600\;\AAA) - m(4400\;\AAA)$.  Although most of the $B$--dropout
galaxies have relatively blue rest--frame UV--to--optical colors, the
brightest galaxies in $\ks$ have redder optical--infrared colors with
a fairly well--defined ``blue--envelope''. Because this sample is
selected in the $\acsz$--band, there is no reason to expect that the
lack of bright galaxies ($\ks \lsim 24$) with blue colors is due to
some selection effect.  At fainter magnitudes, part of this trend may
be due to increasing photometric uncertainties.  However, the trend
among the $B$--dropouts is apparent even at bright magnitudes where
the photometric errors are small, and exists for the $U$--dropouts
(see below), where the optical and IR data are much deeper
than the GOODS data.

A similar blue--envelope exists for the colors of galaxies at $z\sim
3$.   The right panel of figure~\ref{fig:cmd} shows the $\wfv - H$
versus $H$--band color--magnitude diagram for $U$--dropout galaxies
from the WFPC2 and NICMOS data of the \hdf\ (Papovich \etal\ 2001; M.\
Dickinson \etal\ \textit{in preparation}) and from the WFPC2 and
VLT/ISAAC data of the \hdfs\ \citep{lab03}, and using the
color--selection of Steidel et al.\ (1999).  The $U$--dropout criteria
identify galaxies with a mean redshift, $\bar{z} \simeq 2.6$, and 50\%
completeness limits of $2.0 \lsim z \lsim 3.3$.  Although the galaxy
samples for the HDF--N and --S are originally detected in NIR data,
our tests using an $I$--band--selected sample for the HDF--N yield
essentially no change in the number of selected $U$--dropout galaxies
nor in the number and luminosity densities derived below.  The
advantage of using these data is that they provide band--matched
catalogs from optical--to--NIR wavelengths that are comparable to the
data for the $B$--dropout galaxies.

For the $z\sim 3$ galaxies, the $\wfv - H$ color corresponds to nearly
identical rest--frame colors as $\acsi - \ks$ for the galaxies at
$z\sim 4$.  The color--magnitude trends at $z\sim 3$ and 4 are
strikingly similar.  Thus, the LBGs at both $z\sim 3$ and $z\sim 4$
seem to require that either (or likely some combination of) the mean
stellar population age, metallicity, and/or dust opacity increase with
optical luminosity.  Furthermore, we also note that the $U$--dropout
galaxies have slightly redder rest--frame $m(1600\AAA) - m(4400\AAA)$
colors as evidenced by the fact that the ``blue--envelope'' shifts by
$\sim 0.2$~mag.  This reddening seems to imply that the mean ages,
metallicity, and/or dust opacity are actually increasing from $z\sim
4$ to 3.


\section{The Integrated SEDs of High--Redshift Galaxies}

To measure the number and luminosity densities, we require the
redshift distribution and effective volumes probed by the $U$-- and
$B$--dropout criteria.  For the $U$--dropout galaxies, we have used
the effective volume derived by \cite{ste99}.  For the $B$--dropout
galaxies, we have simulated artificial LBGs in the GOODS data with a
distribution of colors and sizes that match the observed $B$--dropout
properties (see Ferguson \etal\ 2003).  We then measure photometry for
the simulated galaxies and apply our color--selection criteria to
derive the probability that a LBG with given magnitude and redshift is
detected in the data, $p(m,z)$.  This technique accounts for
incompleteness due to magnitude and surface--brightness effects as
well as the color--selection process.  We then compute the effective
volumes, $V_\mathrm{eff}(m) = \int dz\; p(m,z)\, dV(z)/dz$.  Because
these selection criteria are insensitive to heavily obscured galaxies
and to galaxies with passively evolving, older stellar populations,
our estimates of galaxy densities are strictly lower limits, derived
from the UV--bright population alone.

\begin{figure}
\epsscale{1.0}
\plotone{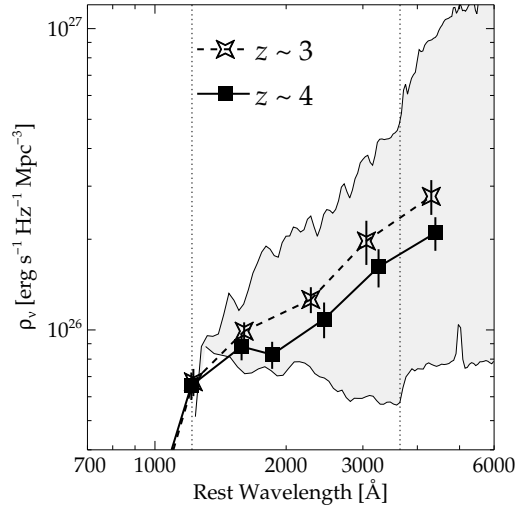}
\epsscale{1.0}
\caption{The co-moving luminosity density as a function of rest--frame
  wavelength.  The data points show the values derived for the
  $B$--dropout galaxies in the GOODS CDF--S data (\textit{filled
  squares, solid lines}) and for the $U$--dropouts in the HDF--N and S
  (\textit{open stars, dashed lines}).  The luminosity densities
  have been integrated down to $m^\ast + 1$~mag with no correction
  for galaxies fainter than this limit or due to dust extinction (see
  text).  The shaded region spans the range of empirical   starburst
  galaxy SEDs from \citet{kin96}, from the bluest [NGC~1705,   $E(B-V)
  \simeq 0.0$] to reddest [$0.6 < E(B-V) < 0.7$] templates.  The
  vertical dotted lines indicate Lyman~$\alpha$ and the Balmer
  Break. \label{fig:sed}}
\end{figure}

To calculate the galaxy densities we consider galaxies brighter than
$m^\ast + 1$~mag, where $m^\ast(\mathcal{R}) = 24.20$ for the
$U$--dropouts (Steidel \etal\ 1999; where $\mathcal{R} \approx
(\wfv+\wfi)/2$, and $m^\ast$ has been adjusted slightly to account for
differences in cosmology and the mean redshifts of the samples here
and those of Steidel et al.), and $m^\ast(I) = 24.70$ for the
$B$--dropouts (Giavalisco \etal\ 2003b).   We derive specific
luminosity densities by integrating the flux densities in each
bandpass to $m^\ast + 1$~mag (where the $B$--dropout galaxies are
well--detected in both the ACS and ISAAC data, see
figure~\ref{fig:cmd}): $\rho_\nu = \int dL_\nu(m)\, n(m)\, L_\nu(m) /
V_\mathrm{eff}(m)$, where $n(m)$ is the observed number density,
$L_\nu(m) = 10^{-0.4(48.6+m)}\, 4\pi d_L^2(1+z)^{-1}$, and $d_L$ is
the luminosity distance.  We plot the measured values in
figure~\ref{fig:sed} along with uncertainties from a bootstrap
resampling.  Note that we have made no correction for fainter galaxies.

The area of the HDFs and the portion of the GOODS field with ISAAC
imaging are fairly small, and cosmic variance may be a substantial
source of uncertainty (see Somerville \etal\ 2003).  To limit this
effect on our results, we have normalized the luminosity densities of
both the $U$-- and $B$--dropout samples such that they match the
rest-frame UV values of Steidel \etal\ (1999) and Giavalisco \etal\
(2003b), respectively, while preserving the measured ``color'' between
the rest--frame UV and other wavelengths.  These surveys average over
more area and sightlines than considered here, which significantly
reduces the uncertainties from cosmic variance.  Cosmic variance may
still be a factor for the $B$--band luminosity densities. The
luminosity weighted mean rest-frame $m(1600\,\AAA)-m(4400\,\AAA)$
colors of the $U$--dropout galaxies in the HDF--N and -S vary by
$\simeq 0.15$~mag, which provides an estimate on the uncertainty in
the typical luminosity--density colors of LBGs in single HDF-sized
fields.  Future studies using the entire GOODS dataset (with full
$U$--band and NIR observations of both fields) will substantially
increase the sample size of both the $U$-- and $B$--dropouts by
roughly a factor of $\approx 25$ and 10, respectively, and should
strengthen our results.

The fact that the rest--frame 1600~\AA\ luminosity density is roughly
unchanged (to within $\lsim 10$\%) implies that the star--formation
rate (SFR; at least of massive, OB stars) is roughly  constant (see,
\eg, Madau, Pozzetti, \& Dickinson 1998).  However, we observe
marginal evidence for a steepening  of the UV continuum ($\sim
1500-2500$~\AA), and strong evidence ($\approx 98$\% confidence) that
the average $m(1600\AAA) - m(4400\AAA)$ color becomes redder for the
$U$--dropouts relative to those for the $B$--dropouts.  In particular,
the luminosity density at rest--frame 4400~\AA\ grows from
$2.1\pm0.3\times 10^{26}$~erg s$^{-1}$ Hz$^{-1}$ Mpc$^{-3}$ at $z\sim
4$ to $2.8\pm 0.3\times 10^{26}$~erg s$^{-1}$ Hz$^{-1}$ Mpc$^{-3}$ at
$z\sim 3$.  Based on the observed color distribution of LBGs (to
fainter magnitude limits than our samples here), our simulations
suggest that this result is unlikely to be biased significantly by the
color--selection of these galaxies.  This change in the mean color
suggests that the mean stellar mass--to--light ratio of these galaxies
is also increasing.  Therefore, the total stellar mass density of this
population is presumably growing by more than the $\gsim 33$\%
increase seen in the rest--frame $B$--band optical light.
\citet{dic03} found evidence for a substantial build--up in the global
stellar mass density from $z\sim 3$ to $z\sim 1$.  The present results
extend that trend to still higher redshifts.

The luminosity densities constrain the global, average
star--formation histories of galaxies at $z\sim 3-4$.   One possible
scenario is a roughly constant SFR with an unchanging dust content.
This would produce approximately equal $\rho_\nu(1600\AAA)$ at $z\sim
3$ and 4, and build up the global stellar mass with time, which would
redden the average galaxy SED from $z\sim 4$ to $z\sim 3$ as the
$B$--band light increases.  We note that the Universe ages by $\sim
50$\% from $z\sim 4$ to 3, and may imply that galaxies formed stars at
a near--constant rate from $z\gg 6$ to 3.  An alternative
scenario is a rising SFR from $z\sim 4$ to 3 with an
increasing dust content that is tuned to produce a roughly constant
$\rho_\nu(1600\AAA)$, and is consistent with the suggestion that the
UV continuum is redder at $z\sim 3$. The fractional stellar--mass
build-up in this scenario would be still greater
than that with constant SFR and an unchanging dust content.

Both of the above scenarios invoke a SFR that is either constant or
growing in time.   However, the stellar populations observed in $z\sim 3$
galaxies appear to be young (a few $\times 10^8$~yr for simple,
monotonic star--formation histories; Sawicki \& Yee 1998; Papovich
\etal\ 2001; Shapley \etal\ 2001), which implies that these stars had
not formed by $z\sim 4$.  Galaxies at these redshifts probably
have star--formation histories that are more complex (with multiple
formation episodes) than those described by the simple models (see
Ferguson \etal\ 2002).  However, even though the star--formation
histories for individual galaxies may be fairly stochastic, the
global, average SFR at high redshift must be roughly constant in order
to explain the evolution in the UV and optical luminosity densities.

Although the increase in the observed optical luminosity density from
$z \sim 4$ to $z\sim 3$ appears robust, the uncertainties due to
cosmic variance and other systematics are non-negligible.  If there
is, in fact, no evolution in the optical luminosity density then this
could imply that stars formed in the $B$--dropout galaxies are not
present within the $U$--dropout galaxies, and that at some point the
$z\sim 4$ galaxies cease forming stars and  evolve beyond the
$U$--dropout selection criteria.  This would have interesting
implications for constraints on the sources of reionization and/or on
the IMF at higher redshifts (see, \eg, Ferguson \etal\ 2002).
However, this would likely be inconsistent with the low number density
of red, $z\gsim 2$ objects \citep{fra03}.  Moreover, the majority of
galaxy evolution models predict growth in the stellar--mass density by
amounts consistent with our observations (although specific
predictions vary from model to model; see Dickinson \etal\ 2003, and
references within), and this suggests that much of the stellar--mass
assembly at high--redshifts does occur in these UV--bright galaxies.


\section{Summary}

We have compared the rest--frame UV--optical colors of color--selected
galaxies at $z\sim 4$ to those of similarly selected galaxies at
$z\sim 3$. We find a great degree of similarity in the rest--frame
$m(\mathrm{UV}) - B$ versus $B$ diagram for galaxies from $z\sim 3-4$.
However, although the rest--frame UV luminosity densities at $z\sim 4$
and $z\sim 3$ are comparable, there is evidence that the rest--frame
$B$--band luminosity density grows by $\approx 33 \pm 16$\%.  Even
though the star--formation histories of individual galaxies may
involve complex and stochastic processes, the evolution of the
luminosity density corresponds to a globally average SFR that is
constant or increases with time.  This implies that the average
stellar-mass--to--light ratio of galaxies is also increasing over
this redshift range and that the global stellar--mass density grows by
more than $\gsim 33$\% over the $\sim 1$~Gyr interval that elapses
between $z\sim 4$ and 3.

By selecting in the rest--frame UV, we are likely to miss galaxies
without ongoing, relatively unobscured star formation.  For example,
\citet{fra03} have used deep NIR images of the HDF--S to identify red
galaxies which may have evolved, massive stellar populations at $z\sim
2$, and may contribute significantly to the global stellar mass
density.  Such objects would be missing from the samples here, leading
to an underestimate of the total density.  Logically, the number and
mass content of these red galaxies should grow as the Universe ages.
Therefore, the apparent increase in stellar mass density from $z\sim
4$ to 3, as traced here by the UV--bright population, would only be
strengthened if red, UV--faint objects were also considered.
Ultimately, the full GOODS dataset will permit more direct tests of
the star--formation histories of the distant galaxy population,
as will future observations with the \textit{Space Infrared
Telescope Facility} and the \textit{James Webb Space Telescope}.

\acknowledgements 

We wish to thank our colleagues for stimulating conversations, and the entire
GOODS team for their concerted effort.  Support was provided by NASA
through grant GO09583.01-96A from STScI, which is operated by the
AURA, under NASA contract NAS 5-26555.  Support for this work, part of
the {\it SIRTF} Legacy Science Program, was provided by NASA through
Contract Number 1224666 issued by the Jet Propulsion Laboratory,
under NASA contract 1407. PM acknowledges support by NASA through
grant NAG5-11513.


\end{document}